# Scalability in Simulating a Large-Aperture, Fresnel Zone Plate Lens for a Conceptual Space Telescope

MANEESHA DUSHMANTHA DE ZOYSA,[1] YANGWOO SEONG,[1] HO XUAN VINH,[2] JAE HUNG HAN,[1] AND HYUN JUNG KIM[1,*]

[1]*Department of Aerospace Engineering, Korea Advanced Institute of Science and Technology (KAIST), 291 Daehak-ro, Yuseong-gu, Daejeon, 34141, South Korea.*
[2]*Institute of Materials Research and Engineering (IMRE), Agency for Science and Technology and Research (A*STAR), 2 Fusionopolis Way, Innovis #08-03, 168634, Singapore.*
*hyunjung.kim1@kaist.ac.kr

**Abstract:** As ambitious space telescope concepts such as ultra-lightweight planar diffractive optical elements (DOEs) emerge, validating the performance remains a major computational challenge. Conventional Fourier propagation algorithms were observed to fail at meter-class apertures due to severe memory limits caused by rigid grid-sampling requirements, and the scaled-down proxy models used for reflector telescopes cannot be applied, since scaling compresses the outermost zones that govern resolution. We benchmarked five Fourier-based propagators against a common Fresnel diffraction integral and found that only those decoupling the focal-plane grid from the input aperture converge within a 1% error threshold. With these findings, we implemented an optimized, stripe-processed Chirp Z-Transform (CZT) framework, evaluating the focal spot strictly within a fixed region of interest to reduce peak memory usage. Applied to five full-aperture configurations from 1.0 m to 5.0 m at f/# = 5, the framework predicted spatial resolution and diffraction efficiency to within 0.001% and 0.16% of analytical references, with modulation transfer function results cross-checked by two analytical extraction methods, all within 6.4 GB of memory on a single consumer-grade GPU. This simulation study represents first steps toward quantifying the expected results of ambitious space telescope concepts and aids the mission development (or selection) phase. With a highly accurate, memory-efficient validation tool, the findings obtained will be used to guide the fabrication decisions of future hardware, optical testing, and physical deployment mechanisms of large-scale diffractive telescopes.

## 1. Introduction

Accurately predicting the optical performance of large-aperture space telescopes through high-fidelity numerical simulations has become a prerequisite for modern sensing missions. Previously, instrument designs and wavefront validations of flagship observatories, such as the James Webb Space Telescope (JWST) and Nancy Grace Roman Space Telescope (RST), have relied on conventional Fast Fourier Transform (FFT)-based methods and matrix propagation libraries (e.g., WebbPSF and PROPER) [1,2]. However, computational limits become evident and stringent when similar techniques are applied to apertures larger than one meter, primarily due to extensive grid-sampling requirements. Circumventing these limits requires physically scaling apertures; for example, simulations of JWST's 6.5-meter primary mirror necessitated a scaled-down 1.08-meter proxy model [1].

Unlike the well-established reflector-type primaries mentioned above, we consider planar diffractive optical elements (DOEs) in this study to be promising for ambitious sensing missions. Among these, the Fresnel Zone Plate (FZP) was selected for its ability to discretely modulate the incident wavefield and yield precise constructive interference at the focal plane by dividing the aperture into concentric arrays of alternating, radially narrowing zones. When etched onto thin structures (e.g., transparent membranes), the device displays promising properties for implementation in spacecraft optics [8,9]. The proposed designs allow for increased aperture size while retaining low areal bulk, thereby inherently overcoming mass, stowage, and rigid surface-figure constraints of mirrors. Using scalable lithographic manufacturing, ultra-lightweight, multi-meter space telescopes can be brought closer to materialization [6,7]. The difficult task of numerically validating optical performance in select DOE geometries is addressed again; however, unlike previous approaches, scaled-down models and subsequent results cannot be referred to. Attempts to scale down FZPs result in the compression of their outermost rings, a faulty change that alters the dimensions of crucial zones responsible for determining the DOE's full-width half maximum (FWHM). Potential membrane deformation patterns (e.g., creases, wrinkles, and ridges) prevent the use of one-dimensional radial methods using azimuthal symmetry. Therefore, two-dimensional, full-aperture wave optics simulations are unavoidable and remain the viable validation technique at large apertures, unless novel solutions suggest otherwise.

In this study, we investigate feasible Fourier-based wave propagation methods to establish a full-aperture, two-dimensional validation framework for large-aperture Fresnel zone plate (FZP) diffractive optical elements (DOEs), without the use of scaled-down geometric models. Five propagation methods are derived from the common Fresnel diffraction integral and developed into functional, MATLAB-based numerical simulations. Namely, the single fast Fourier transform (SFFT), Bluestein's Chirp-Z transform (CZT), the semi-analytical Fourier transform (SAFT), the two-step Fresnel propagator (TSFP), and a hybrid semi-analytical CZT(SA-CZT) are run against constant grid requirements, and a comparative study is done on the accuracy, memory consumption, and runtime results. As a proof-of-principle, we observe that only the CZT and SA-CZT methods converge within the required 1% error threshold for the full-width half-maximum (FWHM), diffraction efficiency (DE), and the modulation transfer function (MTF). We report the following findings specifically within the chosen parameter configuration and the hardware that makes up the testing environment: The study was conducted with monochromatic (λ=532 nm), normally incident, on-axis illumination of a circular binary-amplitude Fresnel zone plate (FZP) in the scalar paraxial regime. In addition, simulation findings were obtained using the following hardware: AMD Ryzen 7 9700X (8c) CPU @3.8GHz, 64GB of DDR5 RAM, and NVIDIA 5060 TI (8 GB VRAM) GPU. Under such conditions, CZT is selected as the superior method for the large-scale study, as it achieves considerable accuracy to SA-CZT while running 14.6 times faster and consuming substantially less memory. We also show that the selected CZT framework, implemented as a GPU-accelerated simulation, scales successfully to meter-class apertures, validating five configurations from D = 1.0 m to D = 5.0 m at f/# = 5 with FWHM error at or below 0.001% and DE error below 0.16%, all within a peak memory budget under 6.4 GB on a single consumer-grade GPU. These results confirm that accurate, full-aperture wave-optics validation of large-scale FZPs is achievable on standard commercial hardware, without the compressed-geometry artifacts introduced by scaled-down proxy models. This framework therefore provides a necessary computational foundation for the future design, fabrication, and optomechanical testing of ultra-lightweight, multi-meter diffractive space telescopes.

## 2. Theoretical Framework

### 2.1 Scalar Diffraction Theory

Maxwell's equations govern the propagation of light, yet at the studied focal ratios ($f\# \geq 5$), coupling between the field vectors near the focus was considered negligible; hence, the field was simplified as a scalar complex amplitude $U(x, y, z)$ [18]. The Rayleigh-Sommerfeld diffraction integral describes propagation from an aperture plane $(u, v)$ to an observation plane $(x, y)$ by distance $z$. By adopting the Fresnel approximation, the obliquity factor is set to unity, the source-to-observation distance is expanded to second order in the exponent as $r \approx z +$ $\frac{[(x-u)^2+(y-v)^2]}{2z}$, and truncated to $r \approx z$ in the denominator [13, 18]. This provides the Fresnel diffraction integral,

$$U(x,y,z) = \frac{e^{ikz}}{i\lambda z} \iint U_0(u,v) \exp\left\{\frac{ik}{2z}[(x-u)^2 + (y-v)^2]\right\} du\, dv \tag{1}$$

Expanding the quadratic term and isolating the observation-plane phase outside the integral rewrites Eq. (1) as a Fourier transform of a chirp-modulated aperture field [13],

$$U(x,y,z) = \frac{e^{ikz}\, e^{\frac{i\pi(x^2+y^2)}{\lambda z}}}{i\lambda z} \iint U_0(u,v)\, e^{\frac{i\pi(u^2+v^2)}{\lambda z}}\, e^{\frac{-2\pi i(xu+yv)}{\lambda z}} du\, dv \tag{2}$$

Equation (2) is the shared target of all five propagators benchmarked in this study; the propagators differ only in how it is discretized [19].

## 2.2 Fresnel Zone Plate Geometry

A FZP achieves focusing by suppressing half-period Fresnel zones, allowing positive, non-zero contributions to arrive at the focus in phase [5]. The exact zone radii are obtained by requiring the path from the $n^{th}$ zone boundary to the focus to exceed the axial path by $\frac{n\lambda}{2}$ [5, 11],

$$r_n^2 = n\lambda f + \frac{n^2\lambda^2}{4}, \quad n = 0, 1, 2, \ldots, N \tag{3}$$

where $f$ represents the designed focal length, and the second quadratic term is the spherical sag correction. When $n\lambda \ll f$, the latter term is considered negligible, and Eq. (3) simplifies to the paraxial form.

$$r_n \approx \sqrt{n\lambda f} \tag{4}$$

Equation (3) is solved for the zone count in all five methods, because, at meter-class apertures, the sag term is no longer small. The radii themselves follow Eq. (4) in the paraxial Fresnel propagators and Eq. (3) in the angular-spectrum propagators. Differentiating Eq. (4) gives a local zone width $\frac{\lambda f}{2}$, and for an aperture of diameter $D = 2r_n$ the outermost zone is

$$\Delta r_N \approx \frac{\lambda f}{D} \tag{5}$$

Equation (5) addresses the resolving power of the device, the finest feature that must be resolved, and links the inverse proportionality to $D$.

## 2.3 Optical Performance Metrics

Considering low numerical aperture conditions, the observed response of the FZP is the Airy pattern of the respective circular aperture. The full width at half-maximum (FWHM) of the function $\left|\frac{2J_1(x)}{X}\right|^2$ is obtained as $1.028(\frac{\lambda f}{D})$, which, by Eq. (5), equals $1.028\Delta r_n$. Until the difference in definition, this is found to be consistent with the Rayleigh form $1.22\Delta r_n$ [11].

Diffraction efficiency (DE) marks the fraction of power incident on the full aperture that reaches the primary order, due to the focusing nature of the DOE. Summing the amplitude contributions of open and opaque zones, uniformly across the entire aperture, gives $\frac{1}{\pi^2} \approx 10.1\%$ for a binary amplitude-type FZP [5]. The amplitude-type case is referenced for the validation stage and considers the $10.1\%$ limit.

The modulation transfer function (MTF) is the modulus, commonly the physical limit, of the optical transfer function (MTF), obtained as the magnitude of the Fourier transform of the simulated point spread function (PSF), normalized to its zero-frequency value [20]. MTF is reported with the FWHM because the zero-order and higher orders of an FZP degrade mid-frequency contrast without broadening the central lobe.

### 2.4 Sampling Limit and Grid Scaling

Equation (2) cannot be calculated in a finite (closed-form) manner for a binary mask; hence, it was numerically assessed on a discrete Cartesian grid. The highest spatial frequency is found at the edge of the aperture, where the zone width is $\Delta r_n$. The Nyquist-Shannon theorem requires at least two samples across that feature, so the input pixel pitch must satisfy [14,18]

$$dx_{\text{in}} \leq \frac{\Delta r_N}{2} = \frac{\lambda f}{2D} \tag{6}$$

and the grid dimension across the aperture obeys

$$N = \frac{D}{dx_{\text{in}}} \geq \frac{2D^2}{\lambda f} \tag{7}$$

At a fixed, known focal length, $N$ grows as $D^2$ and the two-dimensional field as $D^4$. At the fixed f-number ratios, the growth is linear with respect to $D$. The absolute sizes are problematic nonetheless for this study. With respect to the $5.0\ m$ aperture run at $f/\#$=5, Eq. (7) sets $N \gtrsim 4 \times 10^6$, and the grid required to resolve multilevel FZP geometries reaches $N = 1.5 \times 10^7$. Attaining such a complex field within a single matrix array is not feasible; hence, the issue was identified as a constraint, and the study involved comparison of methods for assessing functionality.

### 2.5 Limitations of Conventional Fourier Propagation

The use of a single fast Fourier transform (SFFT) [21] to discretize Eq. (2) pairs the two planes rigidly. The input grid determines the output pitch as

$$dx_{\text{out}} = \frac{\lambda z}{N dx_{in}} \tag{8}$$

so the focal-plane window is set to $\frac{\lambda z}{dx_{in}}$, and the number of samples is locked to N [13,14]. Following Eq. (6) for large $D$ forces $dx_{out}$ far below the scale of the focal spot, and constructing a usable window requires zero-padding by the ratio of the desired window to $\frac{\lambda z}{dx_{in}}$, which multiplies the stored array by the square of that factor [13]. Angular-spectrum propagators (e.g., semi-analytical FT) used in the study preserve the pixel pitch between planes and require comparable padding to suppress wrap-around aliasing [15,22]. Two-step Fresnel propagation gains an adjustable output scale by inserting an intermediate plane, at the cost of a second full-grid transform and a restricted range of intermediate distances [16]. In every case peak memory is set by the padded grid, not by the region examined.

### 2.6 Chirp-Z Transform and Decoupled Sampling

The chirp-z transform (CZT) evaluates the z-transform of a length-$N$ sequence at $M$ points along an arbitrary spiral contour in the complex plane [12],

$$X[m] = \sum_{n=0}^{N-1} x[n]\, A^{-n} W^{mn}, \quad m = 0, 1, \ldots, M-1 \tag{9}$$

Where $A$ is the complex starting point of the contour, and $W$ is the complex ratio between successive points. Setting A = 1, W = exp(−2πi/N), and M = N places the samples at uniform spacing on the unit circle and recovers the discrete Fourier transform, so the CZT is a strict generalization of the SFFT kernel [13]. Matching the discretized kernel of Eq. (2), exp(−2πi $x_m u_n/\lambda z$), to $A^{-n}W^{mn}$ identifies

$$W = \exp\left(-2\pi i \frac{dx_{\mathrm{out}}\, dx_{\mathrm{in}}}{\lambda z}\right), \quad A = \exp\left(2\pi i \frac{x_0\, dx_{\mathrm{in}}}{\lambda z}\right) \tag{10}$$

in which $dx_{out}$, the starting coordinate $x_0$ and the sample count M are free parameters. The output window and its sampling are therefore independent of N, and the field outside the ROI is never formed [13]. Equation (9) is evaluated through Bluestein's identity, $mn = [n^2 + m^2 - (m - n)^2]/2$, which converts it into two chirp multiplications and one linear convolution of length $M + N - 1$ carried out by FFT, giving $O[(M + N)\log(M + N)]$ complexity [12,13]. The kernel is separable, so a two-dimensional propagation is a row pass followed by a column pass, and rows may be transformed in independent batches. Peak memory is then governed by the batch size rather than by $N^2$ [24].

## 3. Simulation Architecture

### 3.1 Simulation Setup

We have developed two separate wave-optics simulation logics using known fast Fourier transform (FFT) methods in MATLAB: one to assess the validity regime of selected aperture diameters and focal-length configurations, and the other to extract the computational resource cost of simulating the validated configuration. A single pipeline logic to handle both is not presented to prevent computational load or their respective remnants from being accounted for during the benchmarking phase.

### 3.2 Single Fast Fourier Transform (SFFT) Method

**Single Fast Fourier Transform (SFFT)**

Paraxial Fresnel propagation by one centred two-dimensional FFT

☐ shared by all five methods ☐ step that differs between methods

**ENTRY: P1_Validate_SFFT(N_in, D, f, $\lambda_0$, cores)**
cores accepted but unused; RAM ceiling enforced pre-flight

↓ tic — timed region opens

timed region → t_exec

**STAGE 1: SAMPLING GRIDS**

**Input plane and analysis window**
dx_in = D / N_in ; N_in forced odd
x_in centred on the optical axis
r_null = 1.22 $\lambda_0$ f / D (Airy null)
MTF window L_MTF = 4 r_null

**Output plane (rigidly coupled)**
dx_out = $\lambda_0$ f / (N_in · dx_in)
= $\lambda_0$ f / D , N_out = N_in
*not a free parameter: one array samples aperture and focal spot*

↓

**STAGE 2: ZONE-PLATE MASK SYNTHESIS**

**Zone map and incident power**
n_exact from $(\lambda_0^2/4)\, n^2 + \lambda_0 f n = R^2$
Total_Zones = ⌊n_exact⌋
pixel open where zone index odd
P_in = pixels(r ≤ R_eff) · dx_in$^2$

**Zone radii and aperture limit**
r_m$^2$ = m $\lambda_0$ f (paraxial radii)
R_eff = min(D/2, $\lambda_0$ f / 2dx_in)
zones finer than Nyquist cut
*radii and kernel both paraxial*

↓

**STAGE 3: PROPAGATION**

**Fresnel-chirped aperture field, one transform**
U_in = mask · exp( i $k_0 r^2$ / 2f ) — chirp written at open pixels only
U_out = fftshift{ fft2[ ifftshift(U_in) ] } · dx_in$^2$ / ($\lambda_0$ f )
*a single fft2 call: no batching, no padding, no explicit parallelism; peak memory ≈ 32 N_in$^2$ bytes on the full input grid*

↓ I(x′, y′) = |U_out|$^2$

**STAGE 4: OPTICAL METRIC EXTRACTION (shared extractor)**

**FWHM of the focal spot**
centre row of I through focus
cubic spline, ×100 upsampling
width between half-maximum crossings either side of peak

**MTF at ν = ν_cut / 2**
crop I to the L_MTF window
OTF = |fft2(I)|, DC-normalised
radial average → read at ν_cut/2
*grid cannot carry ν_cut/2 → NaN*

**Diffraction efficiency of the main lobe**
crop to |x′| ≤ 1.5 r_null , then P_lobe = Σ I(r ≤ r_null) · dx_out$^2$
DE = P_lobe / P_in , P_in = power incident on the full aperture disc
*dx_out = $\lambda_0$ f / D places exactly five output pixels inside r_null at every N_in and every f/# : a structural +6 % bias that grid refinement cannot remove*

↓ toc — t_exec

**RETURN — [ FWHM, DE, MTF, t_exec ] → validation master script**
percent error against FWHM = 1.029 $\lambda_0$ f / D , DE = 0.0840 , MTF_ref

Fig. 1. Simulation logic for the single fast Fourier transform (SFFT) method.

### 3.3 Bluestein's Chirp-Z Transform (CZT) Method

**Bluestein Chirp-Z Transform (CZT)**

The same paraxial Fresnel integral, evaluated on a freely chosen output grid

☐ shared by all five methods ☐ step that differs between methods

**ENTRY: P1_Validate_CZT(N_in, N_out, D, f, $\lambda_0$, cores)**
RAM ceiling enforced pre-flight → OOM before any allocation

↓ tic — timed region opens

**STAGE 1: SAMPLING GRIDS**

**Input plane and analysis window**
dx_in = D / N_in ; N_in forced odd
x_in centred on the optical axis
r_null = 1.22 $\lambda_0$ f / D (Airy null)
MTF window L_MTF = 4 r_null

**Output plane (decoupled ROI)**
L_out = 4 r_null , N_out samples
x_out spans ±L_out/2 evenly
dx_out = L_out / (N_out − 1)
*window and pitch set free of N_in*

↓

**STAGE 2: BINARY-AMPLITUDE ZONE-PLATE MASK**

**Zone map and incident power**
n_exact from $(\lambda_0^2/4)\, n^2 + \lambda_0 f\, n = R^2$
Total_Zones = ⌊n_exact⌋
pixel open where zone index odd
P_in = pixels(r ≤ R_eff) · dx_in$^2$

**Zone radii and aperture limit**
$r_m^2 = m \lambda_0 f$ (paraxial radii)
R_eff = min(D/2, $\lambda_0$ f / 2 dx_in)
zones finer than Nyquist cut
*radii and kernel both paraxial*

↓

**STAGE 3: PROPAGATION**

**Chirp-Z frequency mapping**
u = x_out / ($\lambda_0$ f) , spacing du
W = exp(−i2π · du · dx_in)
A = exp( i2π · u_start · dx_in)
*du · dx_in = 1/N_in gives the DFT*

**Separable batched execution**
row batches of ≤ 250 rows (parfor)
per batch: mask · exp(i$k_0$ r$^2$/2f)
row-wise czt, then cell2mat
column czt · dx_in$^2$ / ($\lambda_0$ f)
*no full N_in$^2$ array is ever formed*

↓ I(x′, y′) = |U_out|$^2$

**STAGE 4: OPTICAL METRIC EXTRACTION (shared extractor)**

**FWHM of the focal spot**
centre row of I through focus
cubic spline, ×100 upsampling
width between half-maximum
crossings either side of peak

**MTF at ν = ν_cut / 2**
crop I to the L_MTF window
OTF = |fft2(I)|, DC-normalised
radial average → read at ν_cut/2
*ROI grid resolves ν_cut/2*

**Diffraction efficiency of the main lobe**
P_lobe = Σ I(r ≤ r_null) · dx_out$^2$ , summed on the decoupled ROI grid
DE = P_lobe / P_in , P_in = power incident on the full aperture disk
*L_out = 4 r_null keeps the whole main lobe inside the integrated window*

timed region → t_exec

↓ toc — t_exec

**RETURN — [ FWHM, DE, MTF, t_exec ] → validation master script**
percent error against FWHM = 1.029 $\lambda_0$ f / D , DE = 0.0840 , MTF_ref

Fig. 2. Simulation logic for the chirp-z transform (CZT) method.

### 3.4 Semi-Analytical Fourier Transform (SAFT) Method

**Semi-Analytical Fourier Transform (SAFT)**

Exact angular-spectrum propagation split by an analytical quadratic fit

shared by all five methods | step that differs between methods

**ENTRY: P1_Validate_SAFT(N_in, N_out, D, f, $\lambda_0$, cores, pad)**
N_out accepted but unused; RAM ceiling enforced pre-flight → OOM

tic — timed region opens

timed region → t_exec

**STAGE 1: SAMPLING GRIDS**

**Input plane and analysis window**
dx_in = D / N_in ; N_in forced odd
x_in centred on the optical axis
r_null = 1.22 $\lambda_0$ f / D (Airy null)
MTF window L_MTF = 4 r_null

**Padded grid and output raster**
N_pad = 4 N_in (odd), same dx_in
zero-pad stops FFT wrap-around
dx_out set by N_pad and the fit
*no free output window: raster is set by N_pad; RAM ≈ 32 N_pad²*

**STAGE 2: BINARY-AMPLITUDE ZONE-PLATE MASK**

**Zone map and incident power**
n_exact from $(\lambda_0^2/4)\, n^2 + \lambda_0 f n = R^2$
Total_Zones = ⌊n_exact⌋
pixel open where zone index odd
P_in = pixels(r ≤ R_eff) · dx_in²

**Zone radii and aperture limit**
$r_m^2 = m \lambda_0 f + (m \lambda_0 / 2)^2$
R_eff = min(D/2, $\lambda_0$ f / 2 dx_in)
mask centred on the padded grid
*exact radii match exact kernel*

**STAGE 3: PROPAGATION**

**Exact angular-spectrum fit**
$k_z(K) = \sqrt{k_0^2 - K^2}$, no paraxial
fit at k_max = 2k₀/3 gives c, d
h(K) = k_z − c k_max − d K²/k_max
*$K^2 > k_0^2$ zeroed (evanescent)*

**Six steps, three full-grid FFTs**
1 fft2 of mask — no input chirp
2 × exp( i h(K) f ), blocked
3–4 ifft2, separable chirp
5–6 fft2 · dx_in², prefactor α

I(x′, y′) = |U_out|²

**STAGE 4: OPTICAL METRIC EXTRACTION (shared extractor)**

**FWHM of the focal spot**
centre row of I through focus
cubic spline, ×100 upsampling
width between half-maximum
crossings either side of peak

**MTF at ν = ν_cut / 2**
crop I to the L_MTF window
OTF = |fft2(I)|, DC-normalised
radial average → read at ν_cut/2
*padded raster resolves ν_cut/2*

**Diffraction efficiency of the main lobe**
crop to |x′| ≤ 1.5 r_null, then P_lobe = Σ I(r ≤ r_null) · dx_out²
DE = P_lobe / P_in , P_in = power incident on the full aperture disk
*P_lobe is further divided by an empirical $(2\pi)^2$ normalisation factor*

toc — t_exec

**RETURN — [ FWHM, DE, MTF, t_exec ] → validation master script**
percent error against FWHM = 1.029 $\lambda_0$ f / D , DE = 0.0840 , MTF_ref

Fig. 3. Simulation logic for the semi-analytical Fourier transform (SAFT) method.

### 3.5 Two-Step Fresnel Propagator (TSFP) Method

**Two-Step Fresnel Propagator (TSFP)**

Paraxial Fresnel propagation relayed through an intermediate plane

shared by all five methods | step that differs between methods

**ENTRY: P1_Validate_TSFP(N_in, N_out, D, f, $\lambda_0$, cores)**
N_out accepted but unused; RAM ceiling enforced pre-flight → OOM

tic — timed region opens

**STAGE 1: SAMPLING GRIDS**

**Input plane and analysis window**
dx_in = D / N_in ; N_in forced odd
x_in centred on the optical axis
r_null = 1.22 $\lambda_0$ f / D (Airy null)
MTF window L_MTF = 4 r_null

**Negative-magnification relay**
dx_out_target = 0.5 µm
m = − dx_out_target / dx_in
$z_1 = f/(1-m)$ , $z_2 = f - z_1$
*m < 0 keeps $z_1$ , $z_2$ > 0 and avoids back-propagation aliasing*

**STAGE 2: BINARY-AMPLITUDE ZONE-PLATE MASK**

**Zone map and incident power**
n_exact from $(\lambda_0^2/4)\,n^2 + \lambda_0 f n = R^2$
Total_Zones = ⌊n_exact⌋
pixel open where zone index odd
P_in = aperture coverage · dx_in²

**Zone radii and aperture limit**
$r_m^2 = m\lambda_0 f + (m\lambda_0/2)^2$
R_eff = min(D/2, $\lambda_0$ f / 2 dx_in)
3×3 sub-pixel sampling per pixel
*greyscale mask curbs aliasing*

**STAGE 3: PROPAGATION**

**Step 1 — source to dummy plane**
U_in = mask · exp( i$k_0 r^2 / 2z_1$)
fft2 · dx_in² → dummy plane
dx_d = $\lambda_0 z_1$ / (N_in dx_in)
*dummy plane sits before focus*

**Step 2 — dummy to focal plane**
chirp $z_1$ , prefactor , chirp $z_2$
fft2 · dx_d² , then output chirp
dx_out = $\lambda_0 z_2$ / (N_in dx_d)
*two FFTs, N_in² grid throughout*

I(x′, y′) = |U_out|²

**STAGE 4: OPTICAL METRIC EXTRACTION (shared extractor)**

**FWHM of the focal spot**
centre row of I through focus
cubic spline, ×100 upsampling
width between half-maximum
crossings either side of peak

**MTF at ν = ν_cut / 2**
crop I to the L_MTF window
OTF = |fft2(I)|, DC-normalised
radial average → read at ν_cut/2
*relay raster resolves ν_cut/2*

**Diffraction efficiency of the main lobe**
crop to |x′| ≤ 1.5 r_null, then P_lobe = Σ I(r ≤ r_null) · dx_out²
DE = P_lobe / P_in , P_in = power incident on the full aperture disk
*P_in is aperture coverage, not the power transmitted through the mask*

toc — t_exec

timed region → t_exec

**RETURN — [ FWHM, DE, MTF, t_exec ] → validation master script**
percent error against FWHM = 1.029 $\lambda_0$ f / D , DE = 0.0840 , MTF_ref

Fig. 4. Simulation logic for the two-step Fresnel propagator (TSFP) method.

### 3.6 Semi-Analytical Chirp-Z Transform (SA-CZT) Method

**Hybrid Semi-Analytical Chirp-Z Transform (SA-CZT)**

SAFT propagation with the final FFT replaced by a CZT zoom

☐ shared by all five methods ☐ step that differs between methods

**ENTRY: P1_Validate_SA_CZT(N_in, N_out, D, f, $\lambda_0$, cores, pad)**
RAM ceiling enforced pre-flight → OOM before any allocation

↓ tic — timed region opens

timed region → t_exec

**STAGE 1: SAMPLING GRIDS**

**Input plane and analysis window**
dx_in = D / N_in ; N_in forced odd
x_in centred on the optical axis
r_null = 1.22 $\lambda_0$ f / D (Airy null)
MTF window L_MTF = 4 r_null

**Padded grid and decoupled ROI**
N_pad = 1.5 N_in (odd)
L_out = 4 r_null, N_out samples
dx_out = L_out / (N_out − 1)
*pad fixes wrap-around only, not output pitch → 7.1 × smaller grid*

↓

**STAGE 2: BINARY-AMPLITUDE ZONE-PLATE MASK**

**Zone map and incident power**
n_exact from $(\lambda_0^2/4)\,n^2 + \lambda_0 f n = R^2$
Total_Zones = ⌊n_exact⌋
pixel open where zone index odd
P_in = pixels(r ≤ R_eff) · dx_in²

**Zone radii and aperture limit**
$r_m^2 = m \lambda_0 f + (m \lambda_0 / 2)^2$
R_eff = min(D/2, $\lambda_0$ f / 2 dx_in)
batched parfor → cell2mat mask
*no input chirp on the mask*

↓

**STAGE 3: PROPAGATION**

**Steps 1–4 — as in SAFT**
$k_z(K) = \sqrt{k_0^2 - K^2}$ exactly
fft2 → × exp( i h(K) f ) → ifft2
separable chirp, rows then cols
*identical to SAFT, at pad 1.5*

**Step 5 — CZT zoom, not fft2**
Δf = dx_out k_max / |4π d_fit f|
row czt then column czt · dx_in²
prefactor α on N_out × N_out
*evaluates only the ROI samples*

↓ $I(x', y') = |U_{out}|^2$

**STAGE 4: OPTICAL METRIC EXTRACTION (shared extractor)**

**FWHM of the focal spot**
centre row of I through focus
cubic spline, ×100 upsampling
width between half-maximum crossings either side of peak

**MTF at ν = ν_cut / 2**
crop I to the L_MTF window
OTF = |fft2(I)|, DC-normalised
radial average → read at ν_cut/2
*ROI grid resolves ν_cut/2*

**Diffraction efficiency of the main lobe**
crop to |x′| ≤ 1.5 r_null, then P_lobe = Σ I(r ≤ r_null) · dx_out²
DE = P_lobe / P_in , P_in = power incident on the full aperture disk
*P_lobe is further divided by an empirical $(2\pi)^2$ normalisation factor*

↓ toc — t_exec

**RETURN — [ FWHM, DE, MTF, t_exec ] → validation master script**
percent error against FWHM = 1.029 $\lambda_0$ f / D , DE = 0.0840 , MTF_ref

Fig. 5. Simulation logic for the semi-analytical chirp-z transform (SA-CZT) method.

## 4. Validation of Studied Methods

### 4.1 Analytical Test for Accuracy of Metrics

We devised a preliminary testing configuration as follows: $D = 0.08\ m, f = 1.5\ m, L = 1.0$, to validate the five studied methods on a binary, negative-centered FZP. Input parameters such as diameter, focal length, and step size (L) were chosen based on ease of fabrication to allow an experimental validation approach on predicted results. The functional test was conducted using each of the five numerical methods; FWHM, DE, and MTF were obtained, then compared against those from the Airy disk intensity pattern obtained mathematically using the first-order Bessel function of the first kind. Hence, we assessed the percent error deviation of each method at increasing grid size for fixed diameter, focal length, and wavelength.

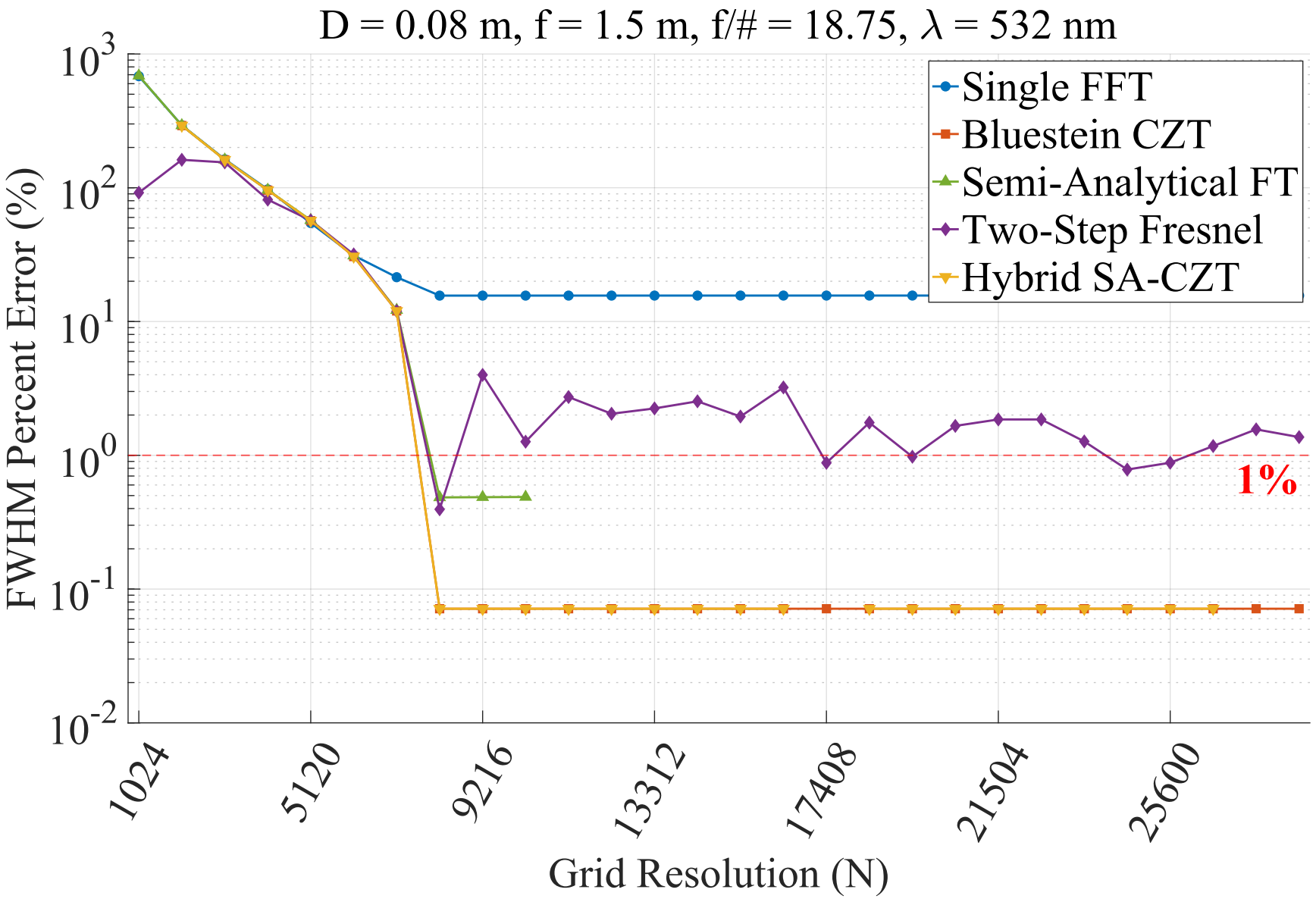

Fig. 6. Simulated percent error of FWHM over varying grid resolution (N).

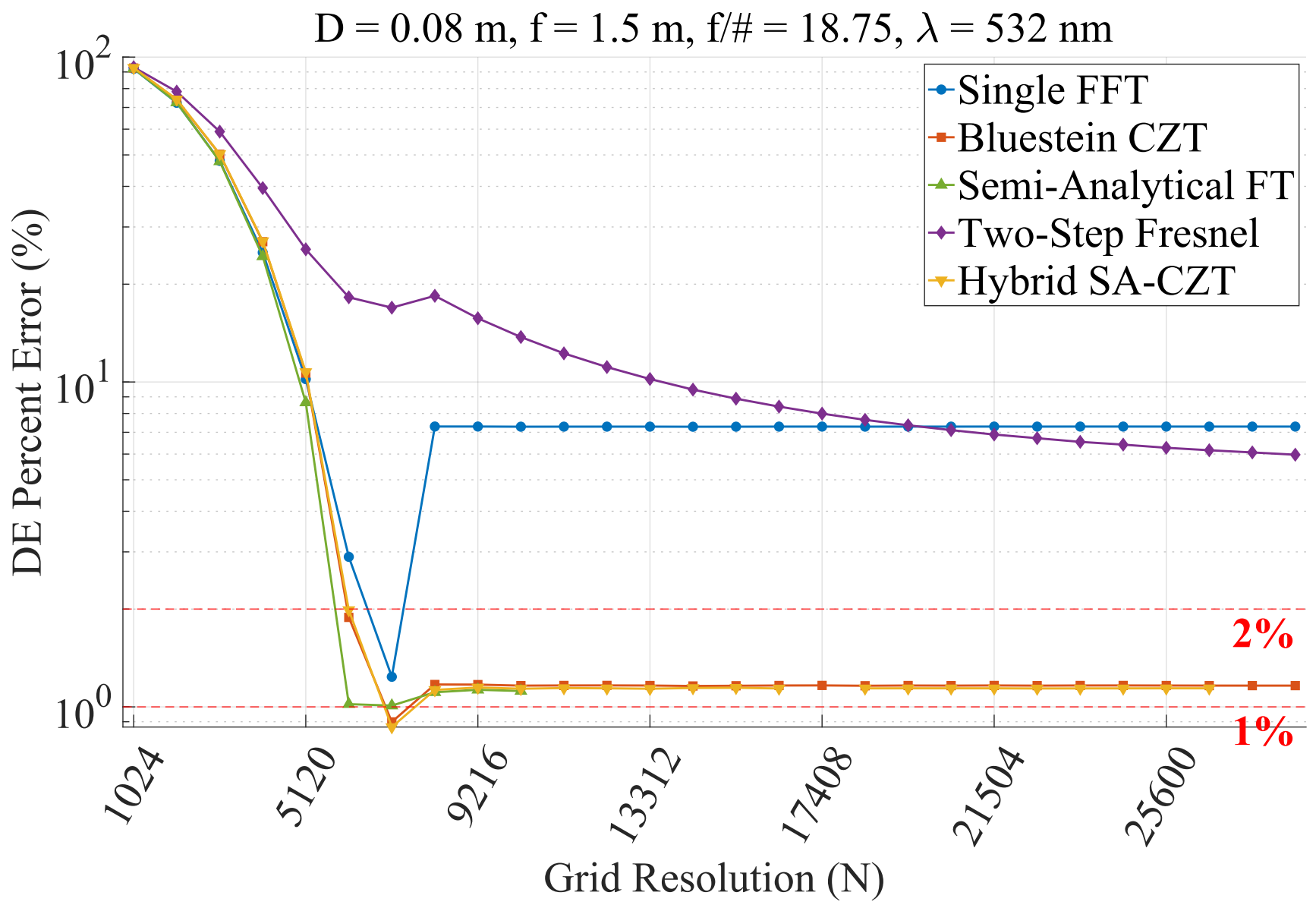

Fig. 7. Simulated percent error of DE over varying grid resolution (N).

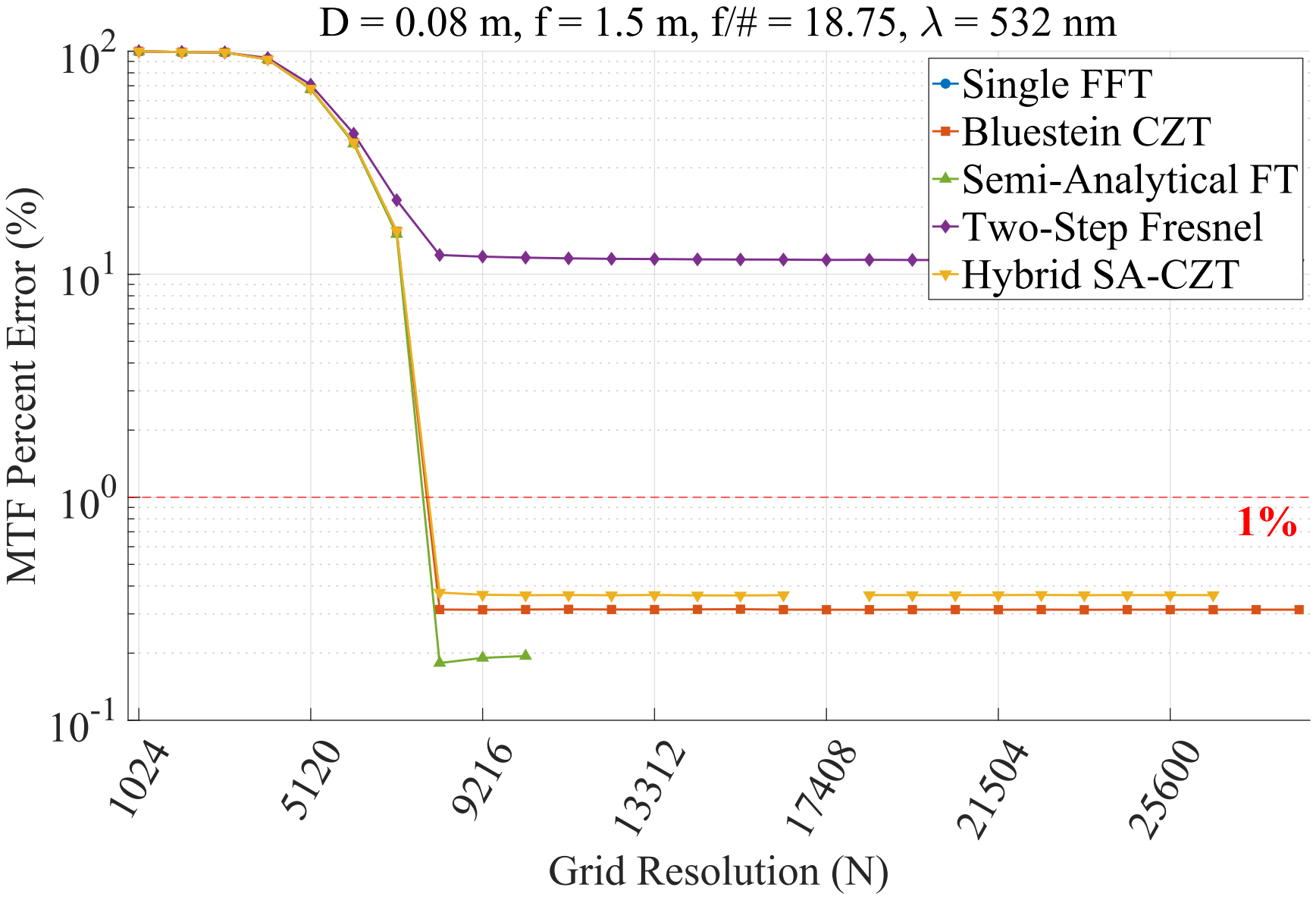

Fig. 8. Simulated percent error of MTF over varying grid resolution (N).

The selected optical performance metrics, namely FWHM, DE, and MTF, can be accurately predicted using the five studied Fourier-based wave-optics methods. We evaluate the degree of accuracy in predicting the desired metrics by obtaining the percent error value per single simulation run, then compare against varying grid sizes to observe potentially converging natures of the trendline. Fig. 6 depicts the simulated accuracy of FWHM predicted by each method under the functional testing parameters. It is observed that all but the TSFP method converge within the chosen grid resolution range. Upon converging, the SAFT method was found to be unable to output results past the grid resolution $N = 11264$ due to out-of-memory (OOM) errors. Of the five, the SFFT method was found to have the highest percent error of 15.6%, a magnitude higher than the remaining methods. With this, simulation methods SFFT, SAFT, and TSFP were considered unfeasible for this study concerned with a large-aperture validation framework for FZPs. Fig. 7 depicts the simulation accuracy of DE under the same parameters. Comparable observations were made: SFFT and TSFP methods result in 7.29% and 5.97% errors, respectively, effectively outside the desired 1% error threshold. The SAFT method was observed to attain a percent error as low as 1.12%, yet it fails past $N = 11264$, whereas CZT and SA-CZT methods maintain percent errors at 1.16% and 1.14% , respectively. From Fig. 8, the trend results were understood to be similar to those of Fig. 7. Importantly, SFFT was not found due to its grid-coupled nature, resulting in its inability to measure MTF. All three figures indicate that of the five methods studied, only two correctly converge at the desired error threshold, and thus support the argument that CZT and SA-CZT remain mathematically accurate and suitable for numerically simulating full-aperture wave optics studies for space telescope concepts. Accordingly, it was understood that the three other methods may not support this scope moving forward as the diameter will be increased.

### 4.2 Experimental Test for Accuracy of Metrics

The delivered accuracy results indicate that we have obtained a series of functioning simulations which are supported analytically by the governing equations of the Fresnel zone plates and the ideal Airy disk diffraction pattern. A separate experimental validation segment was devised due to the need to consider the measurable results of the studied performance metrics post-fabrication.

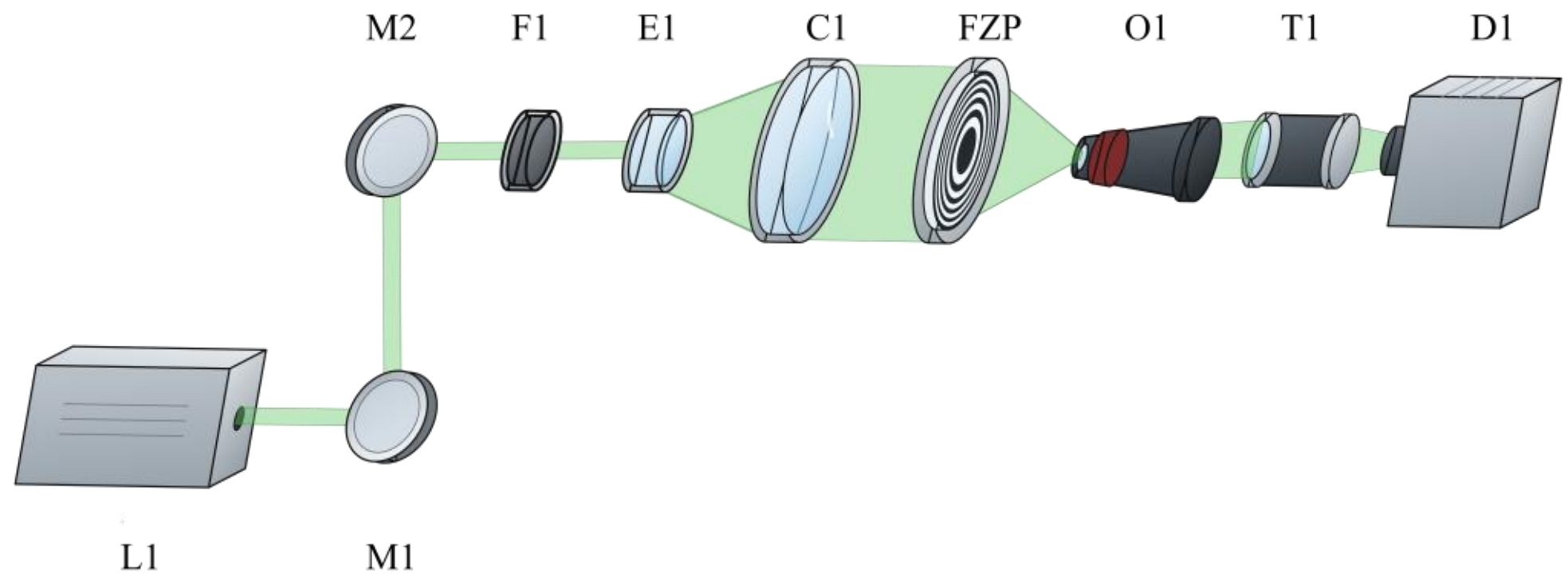

Fig. 9. Schematic diagram of experimental setup used to measure the FWHM, DE, and MTF.

Fig. 9 depicts the apparatus and beam path set up on the optical table to measure the three performance metrics. The ordering of elements (devices) was slightly changed to correctly measure DE and MTF. For DE measurements, we utilized a slim photodiode power sensor (S130C) after the collimation lens (C1) to measure a fraction of the incident power, and at the focal spot for the redirected, focused power. For MTF measurements, a 220-grit diffusion lens followed by a 1951 USAF (positive) resolution target substituted the beam expander (E1). An iris pinhole element was used to restrict unwanted background light, and a Nikon 50x objective lens (O1) was used to precisely capture the diffraction pattern. A chrome, chrome-deposited FZP was fabricated with identical parameters to those of the simulated functional test.

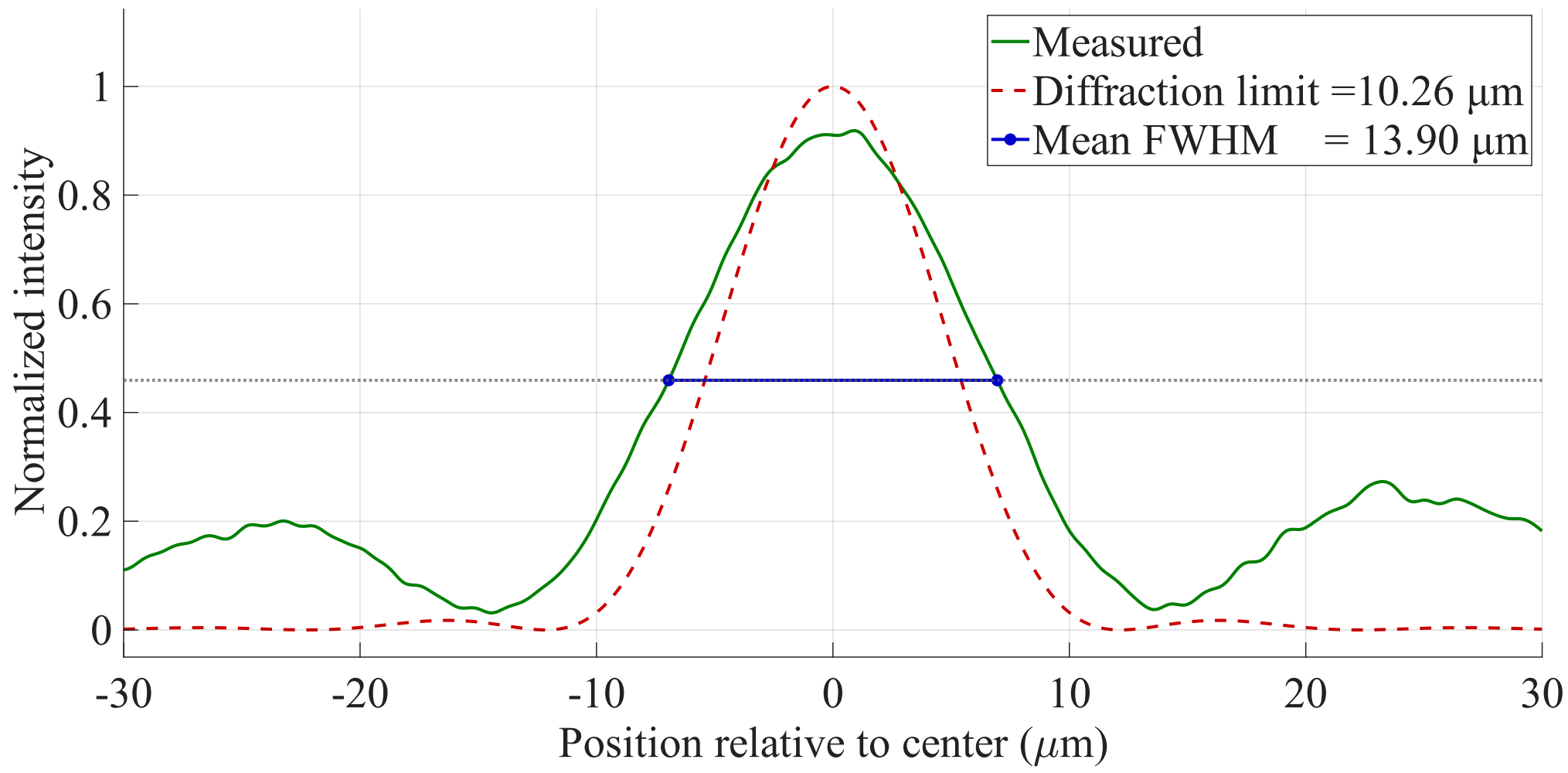

Fig. 10. First summation of radially measured PSF cross-sections.

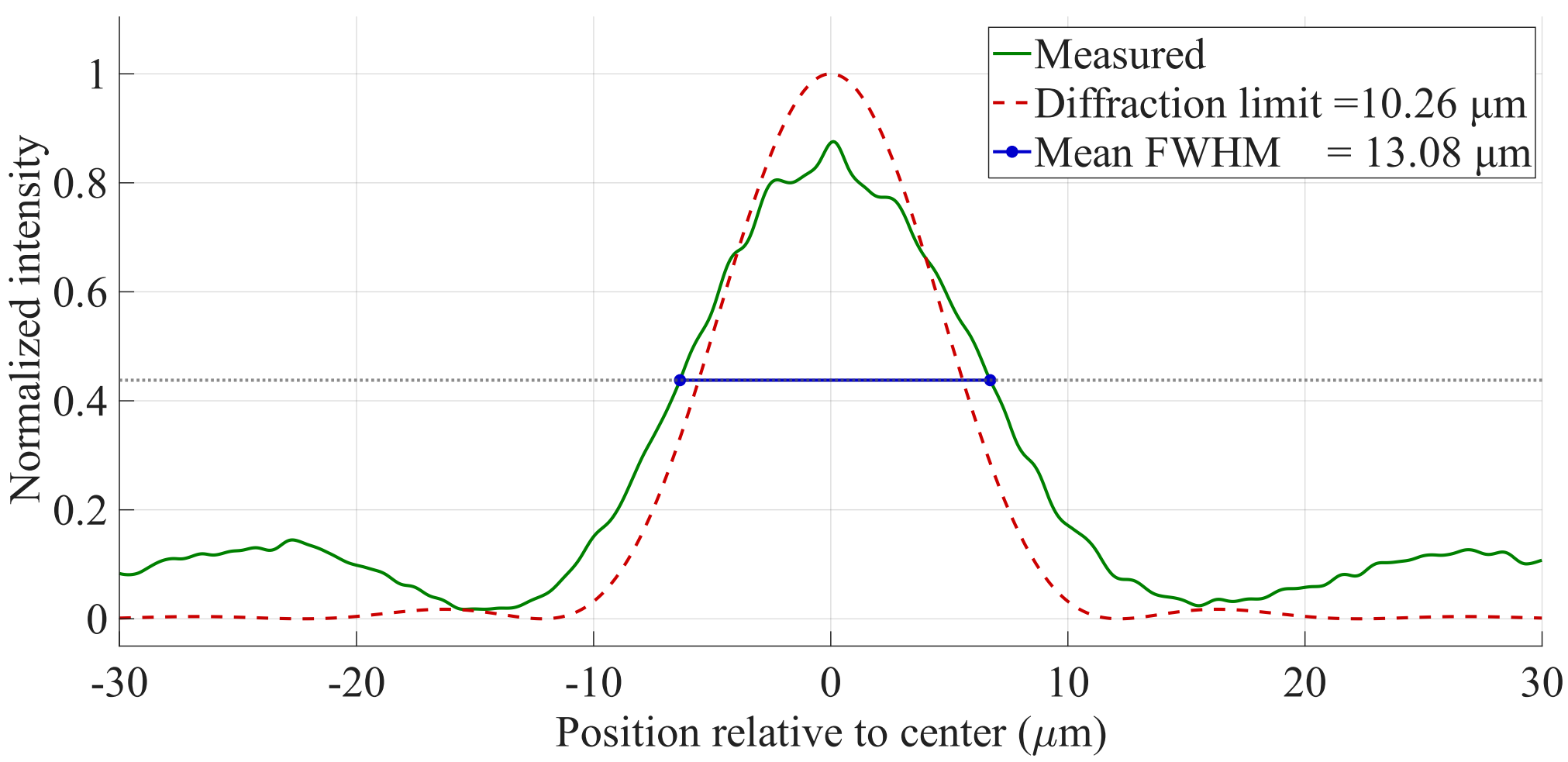

Fig. 11. Second summation of radially measured cross-sections.

The PSF was obtained from the captured focal spot, and aperture-wide radial cross-sections were measured. Contrary to the ideal Airy diffraction pattern, the observed focal spot was found to be unsymmetrical and noisy between the fringe patterns. Multiple trials of radial cross-sections were obtained, specifically 50 trials per plot, from which the FWHM was extracted. An observation was noted that the device's designed focal length $f_{design} = 1500\ mm$ did not match the physically measured focal length $f_{experiment} = 1840\ mm$. Figures 10 and 11 report the measured findings, $FWHM_1 = 13.90\ \mu m$ and $FWHM_2 = 13.08\ \mu m$. A considerable deviation is observed when assessed relative to the literature FWHM.

$$FWHM_1\ percent\ error = \left|\frac{13.90\ \mu m - 10.26\ \mu m}{10.26\ \mu m}\right| \times 100 = 35.48\% \tag{11}$$

$$FWHM_2\ percent\ error = \left|\frac{13.08\ \mu m - 10.26\ \mu m}{10.26\ \mu m}\right| \times 100 = 27.49\% \tag{12}$$

A mismatch in the outermost zone widths of the FZP was initially suspected, as it was the most significant factor contributing to the PSF and focusing. The FZP was inspected under an optical microscope to assess the fabricated zone widths. A deviation of $8.4\% \leq \Delta W \leq 12.1\%$ was found, likely to have originated from the mask aligner's total overlay tolerance.

Consequently, the device-wide systematic errors were quantified; the remaining should be considered due to miscalibration and alignment of the optical setup.

DE was assessed as the ratio of power focused onto the primary order to that incident upon the device aperture. A methodological mismatch was noted as the S130C's sensor area was smaller than the FZP aperture; hence, a ratio of power per surface area was measured, then analytically corrected.

$$Intensity_{focal} = 0.0457\ mW \tag{13}$$

$$Intensity_{incident(S130C)} = 0.0079\ mW \tag{14}$$

$$Intensity_{incident(FZP)} = \frac{0.0079\ mW}{\pi(5\ mm)^2} \times \pi(40\ mm)^2 = 0.506\ mW \tag{15}$$

$$Diffraction\ Efficiency = \frac{0.0457\ mW}{0.506\ mW} \times 100 = 9.04\% \tag{16}$$

$$Diffraction\ Efficiency\ percent\ error = \left|\frac{9.04\% - 10.01\%}{10.01\%}\right| \times 100 = 9.69\% \tag{17}$$

DE measurements were done with an acceptable degree of accuracy, and the remainder of the focused light was likely not collected by S130C due to the size mismatch.

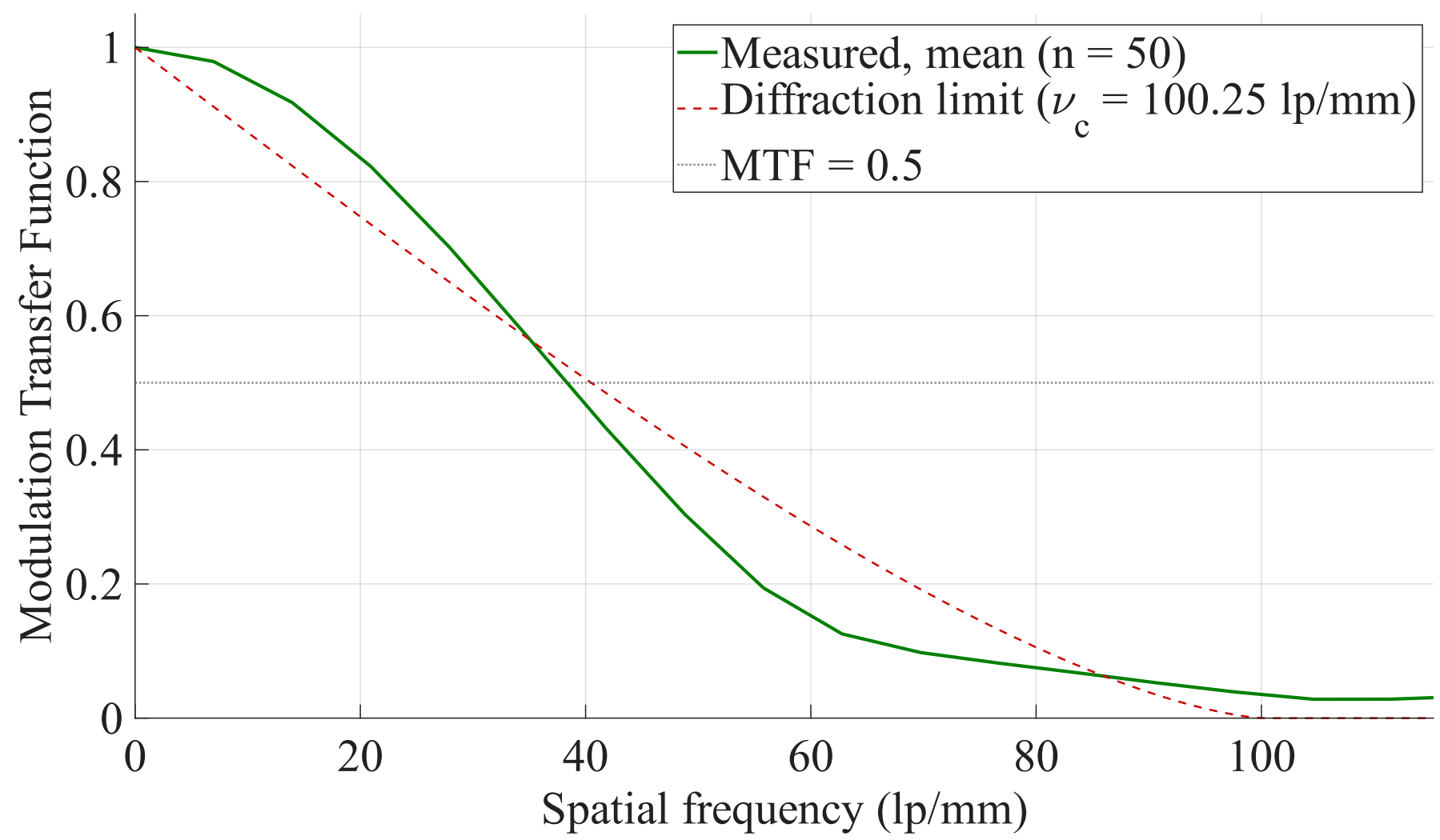

Fig. 12. Experimentally obtained MTF using radial cross-sections of the PSF.

MTF was assessed using two methods: radial cross-sections of the PSF and a 1951 USAF resolution target (the slanted-edge method).

## 5. Scalability and Computational Resource Cost of Methods

### 5.1 Time Consumption as a Cost Per Metric

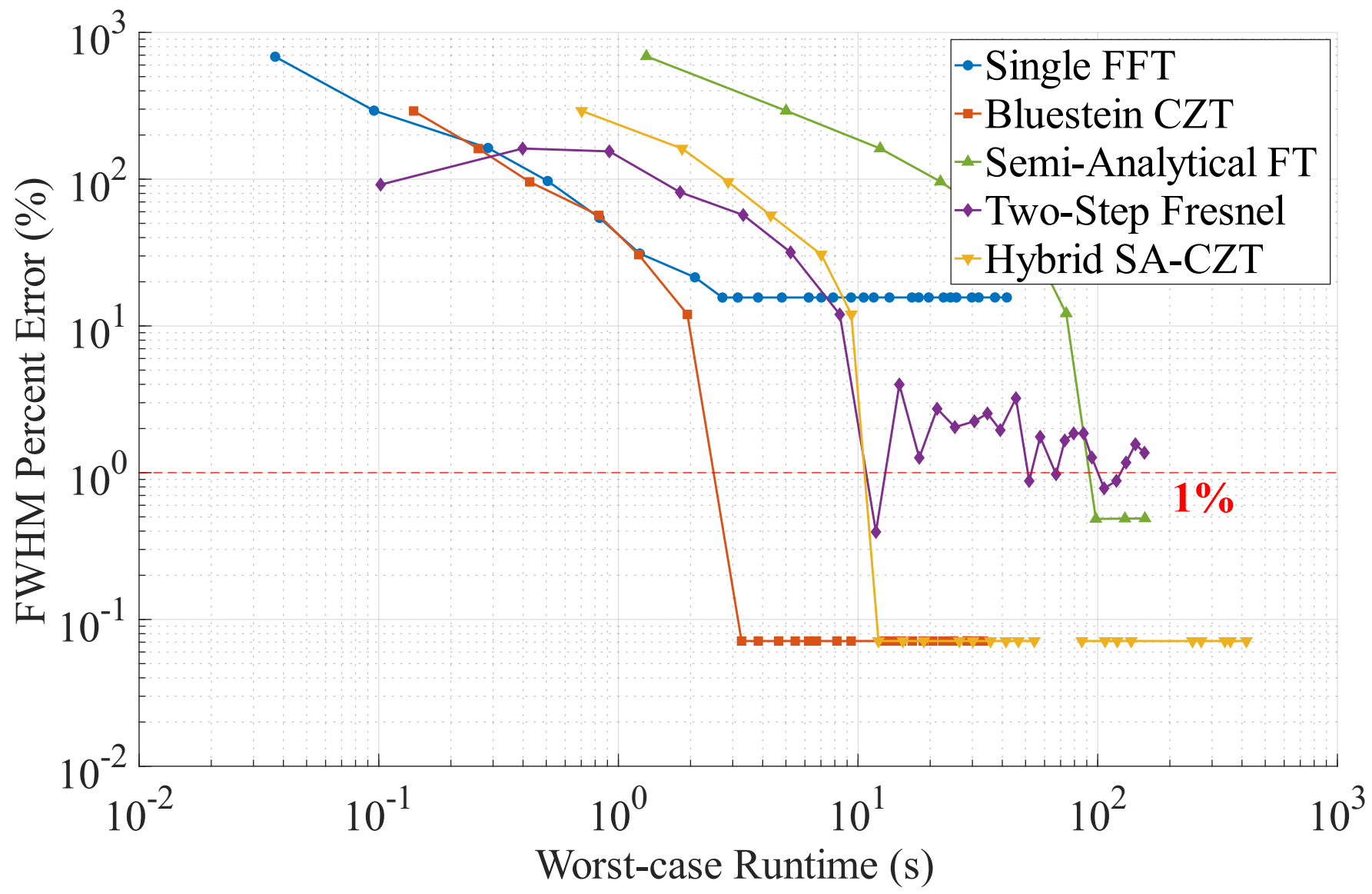

Fig. 13. Simulated FWHM percent error versus simulation runtime

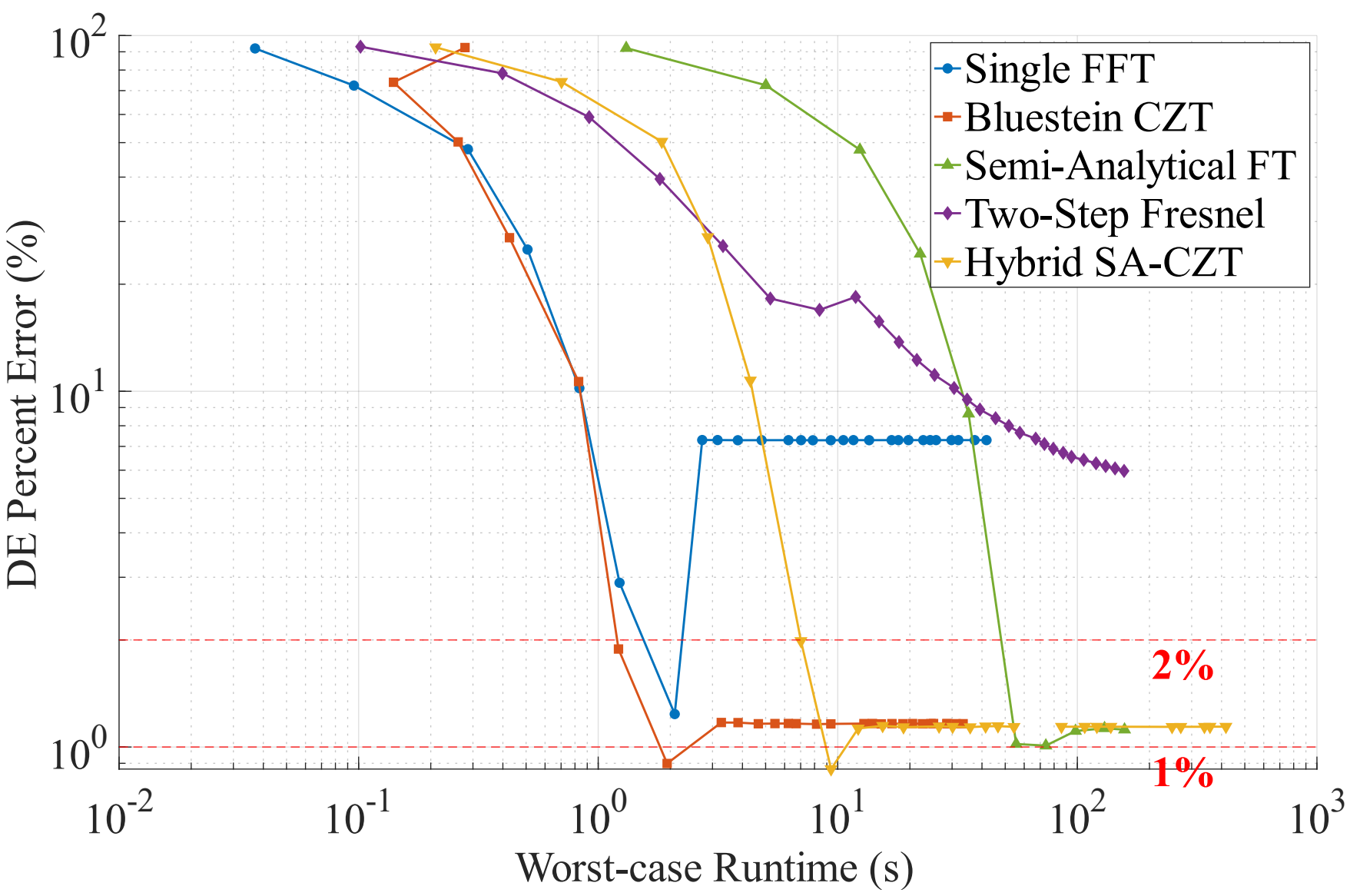

Fig. 14. Simulated DE percent error versus simulation runtime.

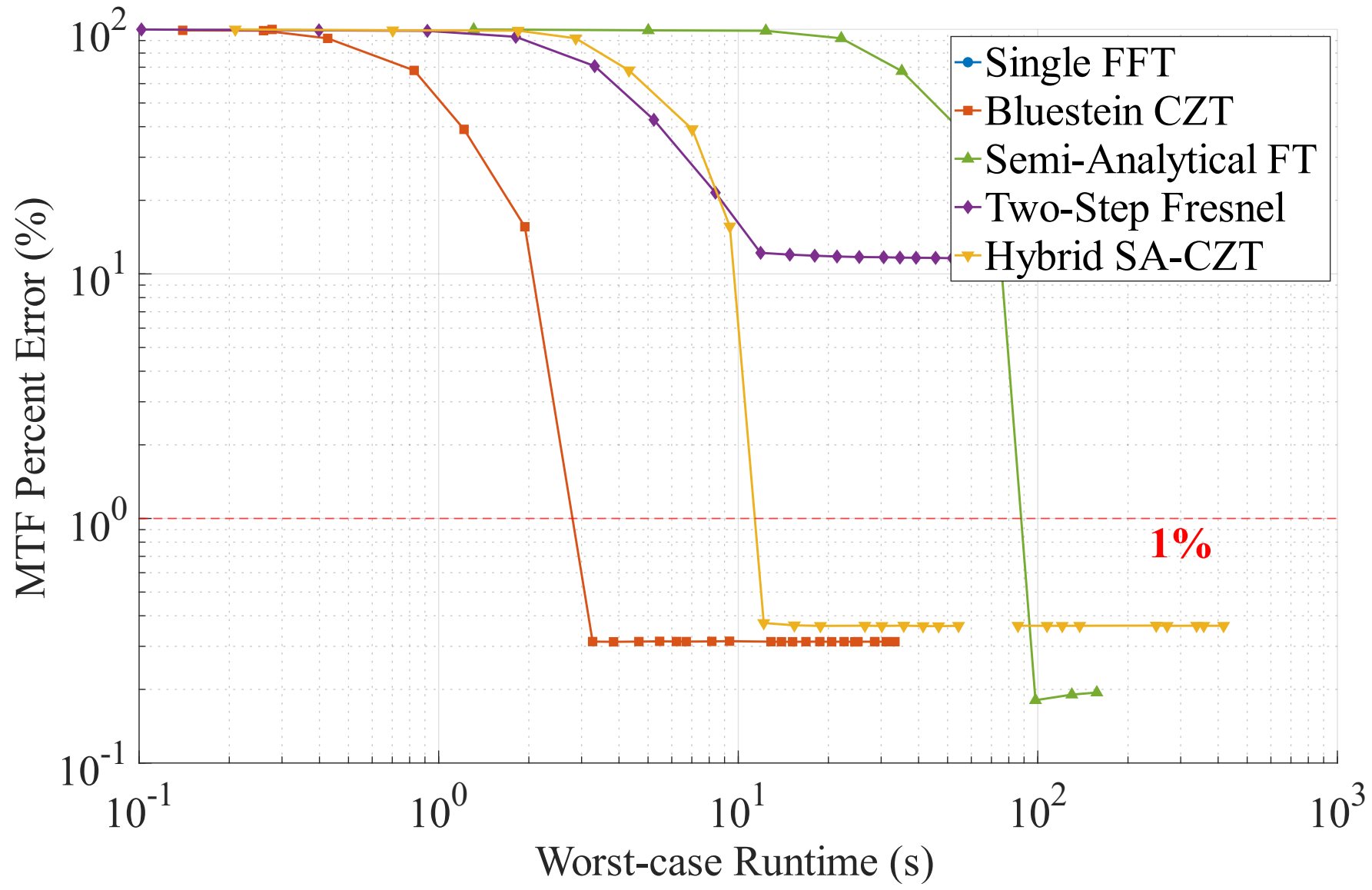

Fig. 15. Simulated MTF versus simulation runtime.

Achieving the necessary degree of accuracy threshold for any simulation must be desirable under realistic runtimes. Time consumption was considered as a computational resource and cost per run. Methods resulting in high runtime were understood to lack optimizations, hence problematic when extended to simulating large-aperture tasks. Figures 9 through 11 depict the metric-wise time cost when attaining accuracies within or near the 1% threshold. Across all three metrics, limitations of SFFT, SAFT, and TSFP are observed and evident in Figures 9 through 11. SFFT fails to achieve the required accuracy threshold (1%); the SAFT method consistently fails past the grid resolution $N = 11264$, whereas the TSFP method fails to converge within the studied range. With these findings, the three methods cannot be relied on when simulating full-aperture, large-scale simulations since the computational cost alone indicates the need for optimization. The CZT and SA-CZT methods are promising as both were equally accurate in predicting FWHM, but SA-CZT remains 0.02% more accurate in DE, while CZT was more accurate in 0.05% in MTF. However, the CZT method is evidently advantageous as it was observed to run 14.6 times faster than the SA-CZT method.

### 5.2 Trends in Runtime and Memory Scale-Up

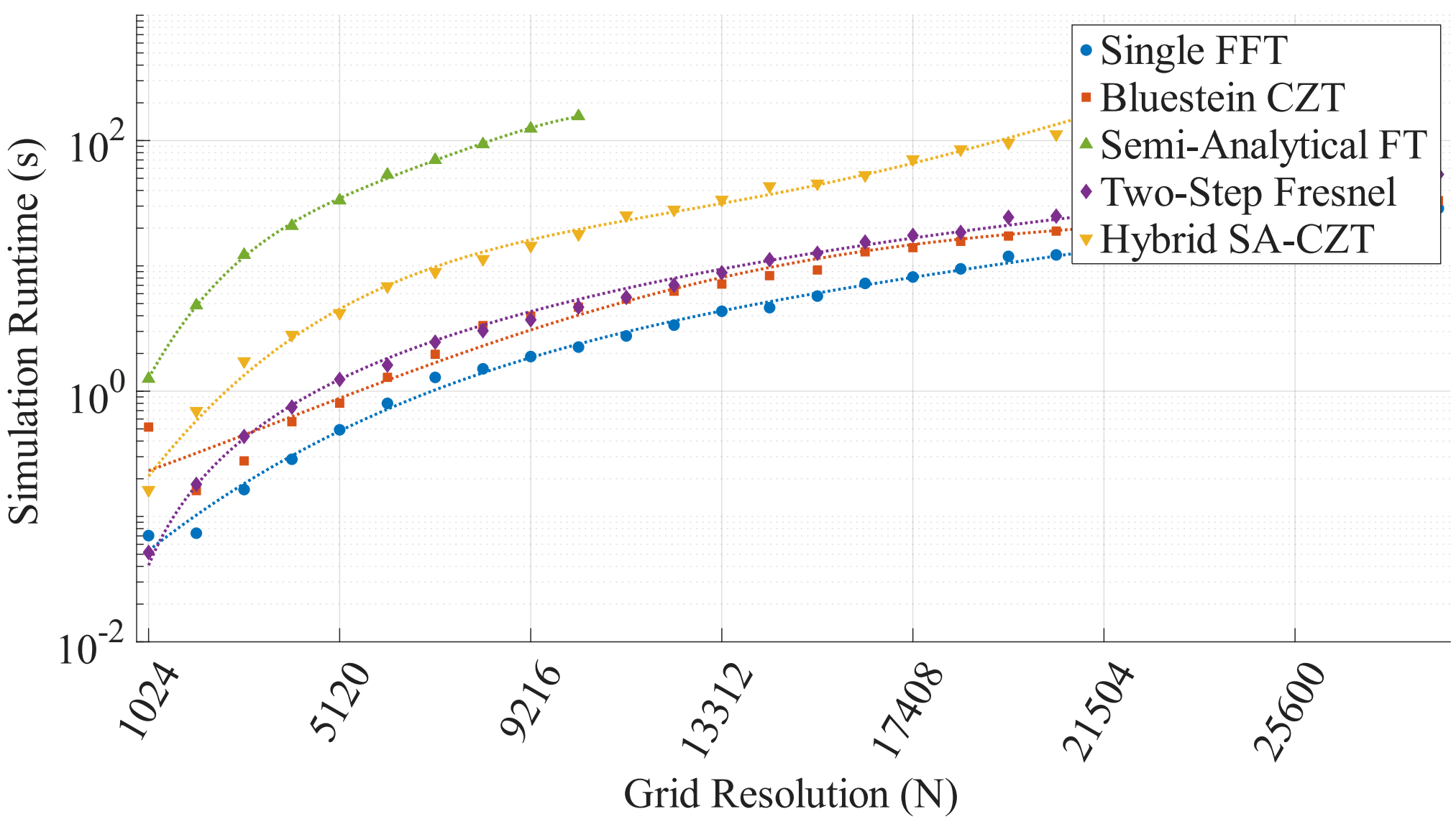

Fig. 16. Simulated runtime scaling per increasing grid resolution (N).

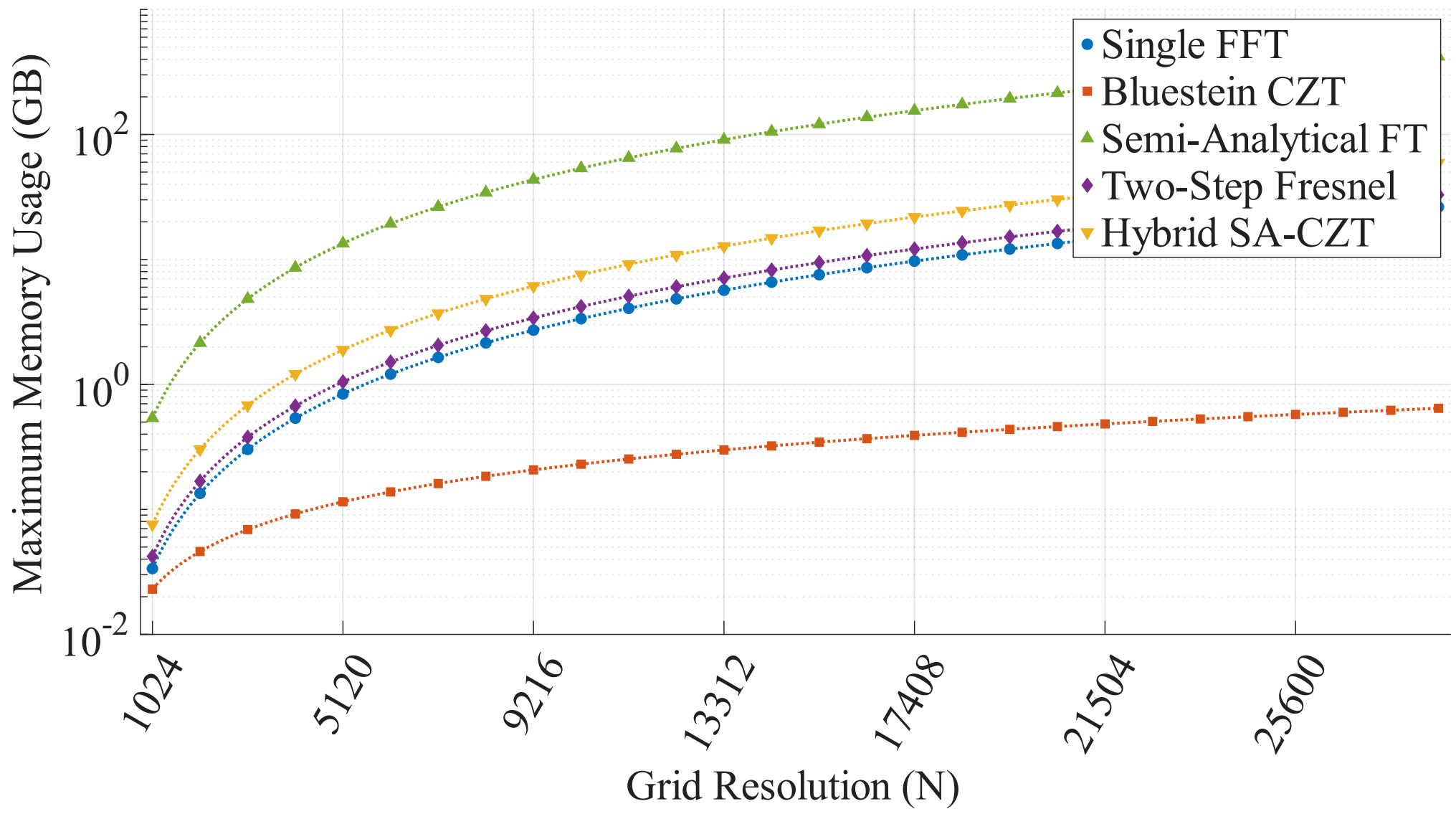

Fig. 17. Maximum memory used versus varying grid resolution (N).

We have studied simulation runtime and memory consumption as dominant, limiting factors of overall algorithmic scalability. Initial considerations were to inspect increasing device aperture; however, it was later decided to resume with the functional configuration ($D = 0.08\,m, f = 1.5\,m, L = 1.0$) for varying grid resolution ($N$). By doing so, we were able to assess the experimentally verified results consistently and tackle scalability as a simulation efficiency problem. Accuracy in predicting the optical performance metrics for larger device apertures relies solely on computationally building finer grid resolutions; hence, any of the studied methods with the highest simulation efficiency, that is, shorter simulation runtimes and smaller memory consumption, will dictate this scalability study. Figures 12 and 13 depict simulation runtime and memory consumption, respectively, of the five studied methods against varying input grid resolution ($N$), ranging from $N = 1024$ to $N = 28672$. The method with the least runtime was observed to be the SFFT method; however, it was previously established that said method does not meet the minimum accuracy threshold. We observe that the CZT method costs the second-lowest runtime while being significantly the lowest in memory consumption. The gradual progressive trendline observed in Fig. 13 of the CZT method outlines its moderate use of computational resources and sustainable nature in algorithmic efficiency.

### 5.3 Method Selection for Large-Aperture Study

We conducted this study to discern the most efficient propagation method, considered suitable for the large-aperture DOE-based space telescope study. From the series of accuracy tests, Bluestein's CZT method was found to produce accurate predictions in the optical performance metrics, mostly within the required threshold. Second to the CZT method, the SA-CZT method was found to deliver nearly identical accuracy in results, albeit being quite similar in structure. The remaining methods were found to be unviable based on the degree of accuracy observed during the functional test. Following this, Bluestein's CZT method was found to attain significantly shorter runtimes and the least memory consumption when benchmarked for increasing grid resolutions. Consequently, we report the selection of the CZT method for simulating large-aperture Fresnel zone plate lenses for the following findings.

a. The CZT method achieves the three metrics prediction with acceptable percent error margins (FWHM, MTF $\leq 1\%$ and DE $\leq 1.5\%$) with a fraction of memory usage (0.644 GB), whereas SA-CZT converges to the studied limit (59.192 GB), $N = 28672$.

b. The CZT method delivers the above accepted accuracies within a fraction of the runtime (33.133s), whereas the SA-CZT method performs much slower (436.01s) at $N = 28672$.

The selection rationale was considered valid and appropriate within the strict bounds and requirements predetermined by the scope of this study. This comparison of methods was tested under the following strict considerations: $D = 0.08\,m$, $f = 1.50\,m$ and $L = 1$ (amplitude-type), the scalar paraxial diffraction regime, an on-axis focal spot with FWHM, DE, and MTF metrics prediction. For such conditions, the error range $1.0\% \leq p.error \leq 1.5\%$ was chosen as acceptable for the task, compared against the ideal Airy diffraction pattern (Bessel function, $1^{st}$ kind).

## 6. Revised CZT Simulation for Large-Apertures

### Full-Aperture MLFZP Simulation Engine

Paraxial Fresnel propagation by GPU-resident separable Bluestein CZT

☐ CPU-side / optical physics ☐ GPU-resident or streamed step

**SCRIPT: mlfzp_fourier_czt_v8.m**
runtime prompts set the FZP level L and the centre configuration

↓ geometry and resources resolved

**STAGE 1: SAMPLING AND OUTPUT GRIDS**

**Grid derived from the element**
Λ_edge = $\lambda_0$ f / R (zone period)
dx_in = Λ_edge / (S_samp max (L, 2))
Nx = ⌈D / dx_in⌉ , forced odd

**Output ROI and CZT phasors**
L_out set directly , N_out samples
u = x_out / ($\lambda_0$ f ) , spacing du
w , a follow from du and u_start
*kernels apply these as phase coefficients*

↓

**STAGE 2: GPU RESOURCE MODEL AND CHECKPOINT LAYOUT**

**VRAM model and batch auto-fit**
nfft = 2^nextpow2(Nx + N_out − 1)
peak = fixed + 3.5 · 8 · nfft · b
*b halves and retries on GPU OOM*

**Chunk groups and resume manifest**
GROUP_ROWS rows → one .bin chunk
pre-flight disk check: Nx · N_out · 8 B
*verified groups are skipped on restart*

↓ batch size b and group size fixed

timed stages → t_row , t_col

**STAGE 3: FUSED GPU ROW STAGE**

**One fused kernel per element**
z = r² / ($\lambda_0$ f ) , L-level phase
Fresnel chirp πz + CZT premultiply
*phase in double, stored single*

**Row Bluestein at fixed length nfft**
FY = fft(U_pad) · fv , G = ifft(FY)
row = wwOut · G(Nx : Nx+N_out−1)
*constant buffers reuse the cuFFT plan*

↓ rows written to disk as raw chunks

**STAGE 4: BLOCK-STREAMED COLUMN STAGE**

**Column blocks streamed back**
two contiguous freads per group
premultiply aa , same fv and wwOut
*both stages share w, a and nfft*

**Output chirp and Fresnel prefactor**
Chirp_out = exp( ik₀ r² / 2f )
Pre = exp(ik₀ f ) dx_in² / (i$\lambda_0$ f )
*I = |Pre · Chirp_out · A_col|²*

↓ I(x′, y′) on the N_out × N_out ROI

**STAGE 5: OPTICAL METRIC EXTRACTION**

**FWHM and Strehl ratio**
centre row at ⌈N_out / 2⌉
parabolic peak , spline ×16 crossings
Strehl = I_peak / (π R² / ($\lambda_0$ f ))²

**MTF and diffraction efficiency**
OTF = |fft2(I)| , DC-normalised
200 radial bins → read at ν_cut / 2
DE = Σ I(lobe) · dx_out² / P_incident
*P_incident = π R² , full aperture power*

↓ verification against the analytical references

**RETURN: result_*.mat , report_*.txt , JSON sidecar → run folder**
FWHM vs Airy and vs zone width , DE vs the L-level ladder , Strehl , MTF

Fig. 18. Revised, multi-level simulation logic using the CZT method.

## 7. Results

### 7.1 Analytical Simulation Performance

**Table 1. Simulation Results for 1.0-meter Aperture (f/# = 5, GPU v8)**

| D (m) | f (m) | f/# | L | $dx_{in}$ (µm) | $N_x$ | VRAM (GB) | Status | Runtime (h) |
|---|---|---|---|---|---|---|---|---|
| 1.0 | 5.0 | 5.0 | 16 | 0.083 | 12,030,077 | 5.04 | Pass | 59.1 |

**Table 2. Simulation Results for 2.0-meter Aperture (f/# = 5, GPU v8)**

| D (m) | f (m) | f/# | L | $dx_{in}$ (µm) | $N_x$ | VRAM (GB) | Status | Runtime (h) |
|---|---|---|---|---|---|---|---|---|
| 2.0 | 10.0 | 5.0 | 8 | 0.166 | 12,030,077 | 6.35 | Pass | 105.8 |

**Table 3. Simulation Results for 3.0-meter Aperture (f/# = 5, GPU v8)**

| D (m) | f (m) | f/# | L | $dx_{in}$ (µm) | $N_x$ | VRAM (GB) | Status | Runtime (h) |
|---|---|---|---|---|---|---|---|---|
| 3.0 | 15.0 | 5.0 | 4 | 0.333 | 9,022,557 | 6.34 | Pass | 74.4 |

**Table 4. Simulation Results for 4.0-meter Aperture (f/# = 5, GPU v8)**

| D (m) | f (m) | f/# | L | $dx_{in}$ (µm) | $N_x$ | VRAM (GB) | Status | Runtime (h) |
|---|---|---|---|---|---|---|---|---|
| 4.0 | 20.0 | 5.0 | 4 | 0.333 | 12,030,077 | 6.35 | Pass | 104.5 |

**Table 5. Simulation Results for 5.0-meter Aperture (f/# = 5, GPU v8)**

| D (m) | f (m) | f/# | L | $dx_{in}$ (µm) | $N_x$ | VRAM (GB) | Status | Runtime (h) |
|---|---|---|---|---|---|---|---|---|
| 5.0 | 25.0 | 5.0 | 2 | 0.333 | 15,037,595 | 6.37 | Pass | 119.0 |

### 7.2 Present Limits

Current GPU-accelerated, revised version of the CZT-based multi-level FZP was observed to be capable of simulating for large apertures ($1.0\ m \leq D \leq 5.0\ m$), yet the limiting factor of the simulation performance was found to be runtime. Necessary efforts were made to decrease time consumption when handling large apertures seen in the previous section, without sacrificing accuracy.

In terms of limits, the simulation scripts can handle D = 5.0, f = 25.0 m, and L up to 2. A 16-level FZP is desirable; however, manufacturing constraints must be applied since the current simulation calculates for a uniform level count across the entire aperture. The two optical performance metrics, FWHM and DE, are predicted to an accuracy of ~ 0.01%. Overall, the simulation is capable of predicting very accurate values for both FWHM and DE; however, the time cost should be considered since, for a configuration of D = 5.0 m, f = 25.0, and L = 2, the expected runtime is 119.0 hours. Since row-wise intermediate data matrices are saved per chosen number of batches, using multiple machines to calculate a section of the circular aperture is a solution to save time.

Studies of other configurations have been considered for continuation in future studies. Accurate simulation was found to be possible with the current state of simulation development; however, the time cost of 119 h was also deemed considerable. Running current simulations for 119 h was thought to be less valuable, and future studies once realistic FZP mask properties are implemented (e.g., deformation profile) were considered more valuable. Hence, the larger diameter studies will be continued in order once the development of realistic properties is added to the FZP.

## 8. Applications to Concept

### 8.1 Design Configuration

From the simulated data obtained for the 1.0-meter configuration (D = 1.0 m, f = 5.0 m, L = 1), a design was explored for a CubeSat-based concept spanning 6U to 12U. Further study of the simulated optical performance metrics will yield better understanding of the optomechanical component; however, currently the simulated field containing the amplitude and phase, focal-spot and diffraction efficiency have been converted into spectral coverage, angular resolution, and image field by considering the CMV1200 CMOS image detector (iSIM90-12U payload, demonstrated by MANTIS CubeSat) [17] alongside the proposed FZP lens.

### 8.2 Spectral Coverage

The FZP's focal length is inherently wavelength-dependent ($f \propto 1/\lambda$) [5], so the system remains monochromatic at 532 nm regardless of detector choice. The CMV1200 natively spans roughly 400 ~ 1000 nm. This study considered a preliminary approach for setting up an optical system, hence a corrector was not considered to allow broadband use [3,4].

### 8.3 Angular Resolution

To obtain the angular resolution, the following formula was used to grasp an early understanding of the detector and FZP configuration's resolving power [11].

$$\theta_{FWHM} = \frac{FWHM}{f}; Rayleigh\ Limit\ \theta_R = \frac{1.22\lambda}{D} \quad (18)$$

$$FWHM(validated\ SIM) = 2.7369\ \mu m = 2.7369 \times 10^{-6}\ m \quad (19)$$

$$\theta_{FWHM} = \frac{2.7369 \times 10^{-6}}{5.0} = 5.474 \times 10^{-7}\, rad = 0.547\, \mu rad = 0.1129\, arcsec \tag{20}$$

$$\theta_R = 1.22\, \times \frac{5.32\times 10^{-7}}{1.0} = 6.490 \times 10^{-7}\, rad = 0.649\, \mu rad = 0.1339\, arcsec \tag{21}$$

### 8.4 Ground Sampling Distance

The ground sampling distance (GSD) defines the ground-projected size of a single detector pixel and determines the finest spatial detail the system can record on the Earth's surface [17,25]. Two GSD values are relevant here: the diffraction-limited GSD, set by the resolving power of the optics, and the detector-limited GSD, set by the pixel pitch of the focal-plane array. These two values are not always equal, and the larger of the two governs the realized system GSD.

The diffraction-limited GSD is derived directly from the simulated focal-spot FWHM. At an orbital altitude H = 500 km, the angular FWHM projects onto the ground as:

$$\text{GSD_diff} = (2.737 \times 10^{-6} / 5.0) \times 500{,}000$$

$$\text{GSD_diff} = 5.474 \times 10^{-7} \times 500{,}000$$

$$\text{GSD_diff} = 0.274\ \text{m} \tag{22}$$

### 8.5 Potential Mission Objectives

With the current ground sampling distance calculations, the FZP system is able to resolve exceptionally well. This includes cars, traffic monitoring, and terrain changes [25]. Information regarding contrast is currently missing but will likely be added during future work developments. With the current amplitude-type, binary FZP, this impressive ground sampling distance may not directly translate to high operational performance due to the inherently low DE [5].

## 9. Discussion

This study establishes a practical computational framework for validating the optical performance of meter-class diffractive space telescopes. We demonstrated that conventional Fourier propagation methods fail at large apertures due to severe memory bottlenecks caused by rigid grid-sampling requirements [18,22]. To overcome this, we implemented an optimized, stripe-processed Chirp Z-Transform (CZT) algorithm that decouples the input aperture grid from the focal plane grid. By evaluating the focal spot strictly within a fixed region of interest, the CZT method drastically reduces memory usage while maintaining a sub-1% error margin for both Full-Width at Half Maximum (FWHM) and Diffraction Efficiency (DE). Ultimately, this approach proves that the optical performance of large-scale Fresnel Zone Plates can be accurately predicted using standard commercial hardware, providing a necessary foundation for future space mission designs.

While certain adaptations are made to the pre-existing Bluestein's chirp-z transform (CZT) [12,13], algorithmic novelty was not explored throughout this study. Relevant batch processing and parallel computing strategies are implemented, yet these techniques are commonly demonstrated in algorithmic studies. Instead, the focus was set on developing a full-aperture wave optics simulation pipeline, viable for predicting the FWHM and DE of large-scale DOEs, strictly under configurations for use in a conceptual space telescope. The approach was kept intentionally elementary as initial scalability was assessed in binary amplitude-type FZP geometries, then later extended to large-aperture, multi-level geometries. CZT has been opted for due to its generous handling of very large matrix arrays with the most efficient use of system

RAM, allowing future works to extend to surface masks (e.g., membrane surface profiles and fabrication errors within geometry) [3,10]. At the very beginning, the JWST surrogate model was introduced at roughly one-sixth of the diameter of its true primary mirror. This study identifies that equivalent techniques cannot be utilized in DOEs and extends to produce a validation engine for the specific concept, which is a large-aperture deployable space telescope. The findings of this study do not improve the CZT method nor parallel computing techniques, yet provides a preliminary platform to study DOEs when scaled up. Future work remains regarding bringing the simulation to produce realistic results in terms of factoring non-ideal aberrations, producing a wider set of performance metrics (e.g., Strehl ratio and wavefront error), and simulation results of larger than meter-class apertures.

While this study successfully validates the decoupled-sampling framework for large-aperture FZPs, simulating highly quantified geometries, such as 16-level phase FZPs [8], remains computationally prohibitive due to the uniform spatial resolution currently applied across the entire aperture. Future work will focus on overcoming this limitation by implementing zone-dependent adaptive meshing. By selectively reducing the quantization levels near the densely packed outer edges while maintaining high phase levels near the center, the computational load can be significantly reduced without compromising optical fidelity. Furthermore, migrating the 1D stripe-processed CZT algorithm to GPU-accelerated clusters and cloud computing environments [10] will be explored to further reduce runtime and enable full-scale, multi-level optomechanical validations.

## Back Matter

### Funding

This work was supported by the National Research Foundation of Korea (NRF) grant funded by the Korean government (MSIT) in 2026 (Grant No. RS-2025-02213804, Project Title: In-Space Servicing and Manufacturing Research Center)

### Acknowledgments

### Disclosures.

The authors declare no conflict of interest.

### Data Availability

The data responsible for the results and figures produced within this paper are not publicly available at this time but may be obtained from the authors upon a reasonable request.

### Supplementary Document

See Supplementary Document 1 for supporting materials.

## References

1. M. D. Perrin, R. Soummer, E. M. Elliott, et al., "Simulating point spread functions for the James Webb Space Telescope with WebbPSF," Proc. SPIE 8442, 84423D (2012).
2. J. E. Krist, "PROPER: an optical propagation library for IDL," Proc. SPIE 6675, 66750P (2007).
3. H. Zhang, H. Liu, W. Xu, et al., "Large aperture diffractive optical telescope: a review," Opt. Laser Technol. 130, 106356 (2020).
4. R. A. Hyde, "Eyeglass. 1. Very large aperture diffractive telescopes," Appl. Opt. 38, 4198–4212 (1999).
5. J. Kirz, "Phase zone plates for x rays and the extreme uv," J. Opt. Soc. Am. 64, 301–309 (1974).
6. L. Koechlin, D. Serre, P. Deba, et al., "The Fresnel interferometric imager," Exp. Astron. 23, 379–402 (2009).
7. D. Serre, P. Deba, and L. Koechlin, "Fresnel interferometric imager: ground-based prototype," Appl. Opt. 48, 2811–2820 (2009).

8. J. A. Britten, S. N. Dixit, M. DeBruyckere, et al., "Large-aperture fast multilevel Fresnel zone lenses in glass and ultrathin polymer films for visible and near-infrared imaging applications," Appl. Opt. 53, 2312–2316 (2014).
9. R. Yamada, H. Kishida, T. Takami, et al., "Optical Fresnel zone plate flat lenses made entirely of colored photoresist through an i-line stepper," Light Sci. Appl. 14, 43 (2025).
10. K. Wei, H. Romero, H. Amata, et al., "Large-area fabrication-aware computational diffractive optics," ACM Trans. Graph. 44, 1–17 (2025).
11. T. Liu, Q. Liu, S. Yang, et al., "Investigation of axial and transverse focal spot sizes of Fresnel zone plates," Appl. Opt. 56, 3725–3729 (2017).
12. L. R. Rabiner, R. W. Schafer, and C. M. Rader, "The chirp z-transform algorithm," IEEE Trans. Audio Electroacoust. 17, 86–92 (1969).
13. Y. Hu, Z. Wang, X. Wang, et al., "Efficient full-path optical calculation of scalar and vector diffraction using the Bluestein method," Light Sci. Appl. 9, 119 (2020).
14. D. G. Voelz and M. C. Roggemann, "Digital simulation of scalar optical diffraction: revisiting chirp function sampling criteria and consequences," Appl. Opt. 48, 6132–6142 (2009).
15. D. Asoubar, S. Zhang, F. Wyrowski, et al., "Efficient semi-analytical propagation techniques for electromagnetic fields," J. Opt. Soc. Am. A 31, 591–602 (2014).
16. D. Wang, J. Zhao, F. Zhang, et al., "High-fidelity numerical realization of multiple-step Fresnel propagation for the reconstruction of digital holograms," Appl. Opt. 47, D12–D20 (2008).
17. R. Guzmán, R. López, E. Ocerin Martínez, et al., "A compact multispectral imager for the MANTIS mission 12U CubeSat," Proc. SPIE 11505, 1150507 (2020).
18. W. Zhang, H. Zhang, C. J. R. Sheppard, et al., "Analysis of numerical diffraction calculation methods: from the perspective of phase space optics and the sampling theorem," J. Opt. Soc. Am. A 37, 1748–1766 (2020).
19. A. Garmendía-Martínez, F. M. Muñoz-Pérez, W. D. Furlan, et al., "Comparative study of numerical methods for solving the Fresnel integral in aperiodic diffractive lenses," Mathematics 11, 946 (2023).
20. A. Ünal, "Numerical Fresnel models of Fresnel zone plates for plane wave at angle of incidence," Sci. Rep. 15, 9246 (2025).
21. J. W. Cooley and J. W. Tukey, "An algorithm for the machine calculation of complex Fourier series," Math. Comput. 19, 297–301 (1965).
22. R. Heintzmann, L. Loetgering, and F. Wechsler, "Scalable angular spectrum propagation," Optica 10, 1407–1416 (2023).
23. A. H. Barnett, "Efficient high-order accurate Fresnel diffraction via areal quadrature and the nonuniform fast Fourier transform," J. Astron. Telesc. Instrum. Syst. 7, 021211 (2021).
24. D. H. Bailey, "FFTs in external or hierarchical memory," J. Supercomput. 4, 23–35 (1990).
25. M. H. B. Azami, N. C. Orger, V. H. Schulz, et al., "Design and environmental testing of imaging payload for a 6U CubeSat at low Earth orbit: KITSUNE mission," Front. Space Technol. 3, 1000219 (2022).
26. D. G. Voelz, *Computational Fourier Optics: A MATLAB Tutorial* (SPIE, 2011).
27. D. Mas, J. Garcia, C. Ferreira, et al., "Fast algorithms for free-space diffraction patterns calculation," Opt. Commun. 164, 233–245 (1999).
28. T. Shimobaba, T. Kakue, M. Oikawa, et al., "Nonuniform sampled scalar diffraction calculation using nonuniform fast Fourier transform," Opt. Lett. 38, 5130–5133 (2013).
29. X. Deng, B. Bihari, J. Gan, et al., "Fast algorithm for chirp transforms with zooming-in ability and its applications," J. Opt. Soc. Am. A 17, 762–771 (2000).
30. F. Cao, Y. Zhao, C. Yao, et al., "All diffractive optical element setup for creating and characterizing optical vortices with high topological charges: analytical models and numerical results," Opt. Commun. 495, 127119 (2021).
31. Z. Wang, S. Zhang, O. Baladron-Zorita, et al., "Application of the semi-analytical Fourier transform to electromagnetic modeling," Opt. Express 27, 15335–15350 (2019).
32. Z. Wang, O. Baladron-Zorita, C. Hellmann, et al., "Theory and algorithm of the homeomorphic Fourier transform for optical simulations," Opt. Express 28, 10552–10571 (2020).
33. K. Matsushima and T. Shimobaba, "Band-limited angular spectrum method for numerical simulation of free-space propagation in far and near fields," Opt. Express 17, 19662–19673 (2009).
34. R. P. Muffoletto, J. M. Tyler, and J. E. Tohline, "Shifted Fresnel diffraction for computational holography," Opt. Express 15, 5631–5640 (2007).
35. T. Shimobaba, T. Kakue, N. Okada, et al., "Aliasing-reduced Fresnel diffraction with scale and shift operations," J. Opt. 15, 075405 (2013).
36. R. Soummer, L. Pueyo, A. Sivaramakrishnan, et al., "Fast computation of Lyot-style coronagraph propagation," Opt. Express 15, 15935–15951 (2007).